\documentclass{article}

\usepackage{arxiv}

\usepackage[utf8]{inputenc} 
\usepackage[T1]{fontenc}    
\usepackage{hyperref}       
\usepackage{url}            
\usepackage{booktabs}       
\usepackage{amsfonts}       
\usepackage{nicefrac}       
\usepackage{microtype}      
\usepackage{lipsum}		
\usepackage{graphicx}
\usepackage{natbib}
\usepackage{doi}
\usepackage{amsmath}

\usepackage{listings}

\lstdefinestyle{mystyle}{
	basicstyle=\ttfamily\footnotesize,
	breakatwhitespace=false,         
	breaklines=true,                 
	captionpos=t,                    
	keepspaces=true,                                                
	showspaces=false,                
	showstringspaces=false,
	showtabs=false,                  
	tabsize=3
}
\title{A Python library for nonlinear system identification using Multi-Gene Genetic Programming algorithm}

\date{}	

\author{{Henrique Carvalho de Castro}\thanks{The authors would like to thank the nancial support from the Brazilian agencies Capes, CNPq and FAPEMIG.} \\
	Universidade Federal de Lavras\\
	Minas Gerais, MG, Brazil\\
	\texttt{henriquec.castro@outlook.com} \\
	\And
	{Bruno Henrique Groenner Barbosa} \\
	Universidade Federal de Lavras\\
	Minas Gerais, MG, Brazil\\
	\texttt{brunohb@ufla.com.br} \\
}

\renewcommand{\shorttitle}{A Python library for nonlinear system identification using MGGP}

\hypersetup{
pdftitle={A Python library for nonlinear system identification using Multi-Gene Genetic Programming algorithm},
pdfauthor={Henrique Carvalho de Castro, Bruno Henrique Groenner Barbosa},
pdfkeywords={System identification, NARMAX, MGGP, Nonlinear systems},
}

\begin{document}
\maketitle

\begin{abstract}
	Models can be built directly from input and output data trough a process known as system identification. The Nonlinear AutoRegressive with eXogenous inputs (NARMAX) models are among the most used mathematical representations in the area and has many successful applications on data-driven modeling in different fields. Such models become extremely large when they have high degree of non-linearity and long-term dependencies. Hence, a structure selection process must be performed to make them parsimonious. In the present paper, it is introduced a toolbox in Python that performs the structure selection process using the evolutionary algorithm named Multi-Gene Genetic Programming (MGGP). The toolbox encapsulates basic tools for parameter estimation, simulation and validation, and it allows the users to customize their evaluation function including prior knowledge and constraints in the individual structure (gray-box identification).
\end{abstract}

\keywords{System identification \and NARMAX \and MGGP \and Nonlinear systems}

\section{Motivation and significance}
System identification is the process through which models are built directly from input and output data. The first step in this process is to choose the mathematical representation that will fit the data. In this sense, the NARMAX (Nonlinear Autoregressive Moving Average with eXogenous inputs) models \cite{leontaritis1985input} are of great interest due to their flexibility and representation capabilities. They are composed of lagged input, output and prediction errors.	The key point in working with such models is the selection of the appropriate structure, which must be as simple as possible, but sufficiently complex to capture the dynamics underlying the data \cite{aguirre2009}.

The Forward Regression Orthogonal Estimator (FROE) \cite{billings1989identification} is a standard algorithm to perform the structure selection of NARX models. It is based on the Error Reduction Ratio (ERR) measure, which evaluates how good each single model term is in explaining the output data variance. There are some important disadvantages in the use of this technique: it has limitations for training data with the presence of certain input characteristics \cite{piroddi2003}; and it suffers from the \textit{curse of dimensionality} with the increment of the degree of non-linearity and higher long-term dependencies. To circumvent these problems, one can resort to Evolutionary Algorithms (EA), such as Genetic Algorithms (GA)  \cite{holland1975,goldberg1988}, Genetic Programming (GP) \cite{koza1992} and Multi-Gene Genetic Programming (MGGP) \cite{hinchliffe1996,hinchliffe2001,hinchliffe2003}. Recently, several modeling and forecasting works have been developed with the use of MGGP (as in \cite{ghareeb2013,mehr2017,safari2018,madvar2019}). The algorithm has shown itself to be very flexible and to present good performance.

To the best of the authors' knowledge, the only available toolbox capable of performing MGGP optimization is the GPTIPS \cite{searson2010gptips}, a machine learning platform for MATLAB focused on symbolic regression. Unfortunately, it is restricted to those who own the licenses which hinders contribution from	the community.

This paper presents the \textbf{mggp} package in \textbf{python}. In addition to the MGGP evolutionary algorithm framework, we encapsulate basic methods to work with NARMAX models, i.e. parameter estimation, model simulation and validation methods; so that the user can easily work on the evaluation function to be optimized. 

\section{Software description}

\sloppypar{
	The current toolbox is focused on the structure selection of NARX/NARMAX models using an evolutionary algorithm called MGGP. To structure the adequate framework that is representative of such models, we attempted to develop a well organized Object Oriented program that comprises all necessary tools to perform the task, i.e. model coding (representation), basic parameter estimation, simulation and validation methods,  and the aimed structure selection algorithm.
}
\begin{figure}[t]
	\centering
	\includegraphics[width=0.45\linewidth]{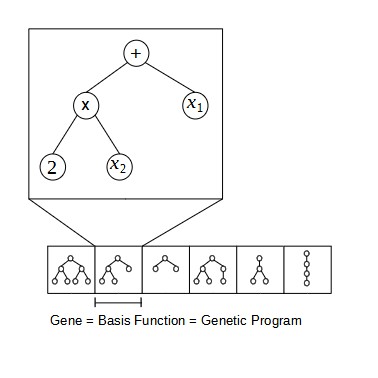}
	\caption{MGGP individual representation. It is a sequel of genetic programs as basis functions. The linear combination of these programs represents the mathematical model. It is highlighted one single gene that represents the mathematical function $2x_2 + x_1$.}
	\label{fig:mggp_representation}
\end{figure}
\subsection{Representation}
In NARMAX models, the current output is obtained from past input-output and residual signals, as follows:
\begin{equation}
	\small
	\label{eq:narx}
	\begin{split}
		y[k]=\mathit{F}^l(&y[k-1],...,y[k-n_y],u[k-1],...,\\
		&u[k-n_u],\xi[k-1]...\xi[k-n_\xi]) + \xi[k],
	\end{split}
\end{equation}
where $F[\cdot]$ is a nonlinear function; $y[k]$, $u[k]$ and $\xi[k]$ are the output signal, input signal and residual vector, respectively; and $n_y$, $n_u$ and $n_\xi$ are their respective maximum lags. These models are extensions of NARX models in which  residual terms are included to remove parameters' bias. In the case of \textit{polynomial} models, the nonlinear function is a polynomial function of degree $l$ ($F^l$). The MGGP fits naturally as a NARMAX description since it can be represented as the linear combination of separated basis functions:
\begin{equation}
	\small
	g(\varphi,\Theta)=\sum^m_{i=1}\theta_ig_i(\varphi),
\end{equation}
where $ m $ is the number of basis functions, $ g_i $ represents individual functions and $ \theta_i $ the model parameters. It can be seen as a GA individual \cite{holland1975,goldberg1988} in which each \textit{gene} contains one GP individual \citep{koza1992,eiben2003,poli2008} as the basis function (see more in \cite{hinchliffe2001} and \cite{hinchliffe2003}). Figure \ref{fig:mggp_representation} graphically presents an MGGP individual representation as a sequel of GP individuals, which are tree representations of mathematical functions. To codify such representation we use the GA and GP frameworks available in the \textit{Distributed Evolutionary Algorithms in Python} (DEAP) \cite{deap2012} package.

\begin{figure}[t]
	\centering
	\includegraphics[width=0.45\linewidth]{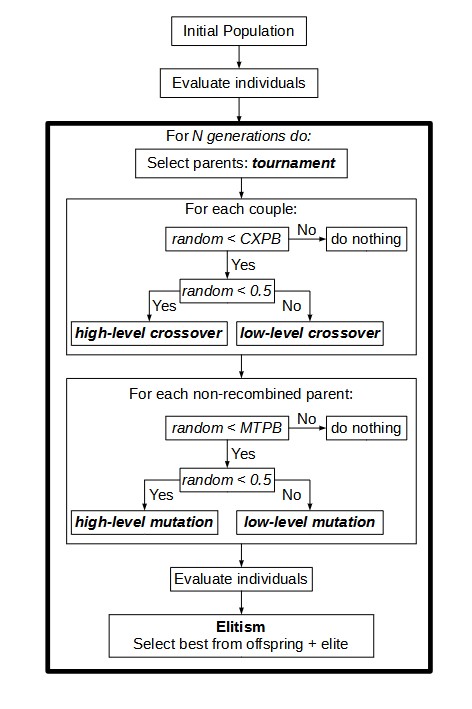}
	\caption{Mono-objective MGGP optimization algorithm flowchart.}
	\label{fig:mggp_flowchart}
\end{figure}
\subsection{MGGP algorithm}
As an EA, the MGGP possesses a standard behavior. It begins with a initial population of random individuals (chromosomes) and at each generation (main loop) the best solutions are sought (\textit{selection}), then combined (\textit{recombination/reproduction/crossover} and \textit{mutation}) in order to generate even better individuals. The MGGP works with two levels of crossover:  \textit{low-level crossover} and \textit{high-level crossover}. In the former, one gene is randomly selected from each parent individual and they exchange GP subtrees as genetic material. In the latter, the genetic materials are exchanged as entire basis functions in a way similar to GA one-point crossover. We include two kinds of mutation in the algorithm: an inner mutation, which occurs as a GP subtree mutation, and an outer mutation, which swaps a gene for a new one, with an entirely new basis function (see \cite{poli2008} for details on GP genetic operators).

Figure \ref{fig:mggp_flowchart} exhibits a flowchart of the toolbox mono-objective MGGP algorithm. It begins with a initial population that is evaluated. Then, the generation loop starts: \textit{i)} the parent individuals are selected via tournament, \textit{ii)} each parent couple has a chance to be recombined (\textit{CXPB}), \textit{iii)} each individual that has not been recombined has a chance to be mutated (\textit{MTPB}), \textit{iv)} the individuals are evaluated, and \textit{v)} the elitism operator is applied. The evaluation function shall be customized by the user.

The MGGP multi-objective evolution uses the NSGA2 framework \cite{Deb2000}. It differs from the mono-objective optimization in the tournament and selection of the next generation (elitism) which must observe the Pareto Optimal Set.

\subsection{Computational implementation}
The \textbf{mggp} package, available on GitHub for pull requests\footnote{https://github.com/CastroHc/MGGP} and in the PyPI repository for installation\footnote{pip install mggp}, is composed of two classes: the \textit{mggpElement class} and the \textit{mggpEvolver class}. 

An \textit{mggpElement} object is responsible for carrying the attributes and methods used to create, simulate and evaluate individuals from an MGGP population. This class is able to build single-input single-output (SISO) and multiple-input single-output (MISO) models and works with NARX and NARMAX representations. It comprises three simulation methods, i.e. one-step-ahead prediction, free-run simulation and multiple-shooting, and their respective mean-squared error (MSE) scores (see \cite{ribeiro_smoothness_2020} for comparison analysis); the Least Squares (LS) and Extended Least Squares (ELS) parameter estimation methods \cite{young1968}; and a standard FROE structure selection method \cite{chen1989orthogonal}. The NARMAX terms, which are GP individuals, are constructed from a set of mathematical functions that can be customized by the user. The \textit{mggpElement} class has a built-in back-shift operator to determine the lag of a term.

An \textit{mggpEvolver} object is responsible to execute the evolution of a population. The individuals from this population are created according to the primitive set of mathematical functions defined in the \textit{mggpElement} object. It is able to perform mono- and multi-objective optimizations.

\section{Illustrative Example}
In this section we illustrate how to use the \textbf{mggp} toolbox through a simple example. We will show (i) how to set up an \textit{mggpElement} object, (ii) how to define a specific model and simulate it, (iii) how to build an evaluation function, and (iv) how to set up an \textit{mggpEvolver} object and run the optimization algorithm.

\subsection{Simulate a system using the \textit{mggpElement} class}
Use the \textit{setPset()} method to define the maximum delay of a single back-shift operator as one ($q1$) and the number of variables as two (one input and one output). This way the primitive set is able to define polynomial NARX models (it is already set to use multiplication function).

\begin{lstlisting}
	from mggp import mggpElement
	element = mggpElement()
	element.setPset(maxDelay=1,numberOfVariables=2)
	element.renameArguments({'ARG0':'y1','ARG1':'u1'})
\end{lstlisting}
Consider a stochastic system given by \cite{piroddi2003}:
\[\small y[k] = 0.75y[k-2] + 0.25u[k-1] - 0.20y[k-2]u[k-1],\]
where $u[k]$ is a White Gaussian Noise with mean of zero and variance one. Use the \textit{createModel()} method to build a model from a list of strings that is representative of the system and then compile it. The model parameters are known.
\begin{lstlisting}
	listStrings = ['q1(y1)','u1','mul(q1(y1),u1)']
	model = element.createModel(listStrings)
	element.compile_model(model)
	theta = np.array([0.75,0.25,-0.20])
\end{lstlisting}
Use the \textit{predict\_freeRun()} method to simulate the system from a initial condition and the input vector.
\begin{lstlisting}	
	u = np.random.normal(loc=0,scale=1,size=(500))
	y0 = np.zeros(2)
	y = element.predict_freeRun(model,theta,y0,u[:-1])
\end{lstlisting}

\subsection{Define the evaluation function and run the optimization}
The MGGP is an optimization algorithm that minimizes an evaluation (or cost) function. In this sense, the \textit{mggpEvolver} class is set to run an evaluation function that receives as \textit{only} argument the individual to be evaluated. We present a very simple example in which we seek to minimize the one-step-ahead MSE and the model parameters are estimated via LS method. An exception treatment must be performed to keep the program running after a Singular Matrix Exception error is raised.
\begin{lstlisting}
	def evaluate(ind):
		try:
			element.compile_model(ind)
			theta = element.ls(ind,y,u)
			return element.score_osa(ind, theta, y, u),
		except np.linalg.LinAlgError:
			return np.inf,
\end{lstlisting}

Use the \textit{mggpEvolver} constructor to set up the MGGP parameters. It must be defined the population size (popSize), the crossover probability (CXPB), the mutation probability (MTPB), the number of generations to be run (n\_gen), the maximum height of the trees that represents model terms (maxHeight), the maximum number of terms an individual may possess (maxTerms), the percentage of individuals from a population to be kept after a generation (elite) and the \textit{mggpElement} object that carries information about individual creation. The evaluation function is sent as an argument of the \textit{run()} method, which returns a log of fitnesses history and the resultant elite population (hall of fame - hof).

\begin{lstlisting}
	from mggp import mggpEvolver
	mggp = mggpEvolver(popSize=500,CXPB=0.9,MTPB=0.1,n_gen=50,
							maxHeight=3,maxTerms=5,verbose=True,elite=10,
							element=element)
	hof,log = mggp.run(evaluate=evaluate)
	model = hof[0]
	
	for term in model:
		print(term)
\end{lstlisting}

\subsection{Multi-objective example}
For a multi-objective optimization, the \textit{mggpElement} must be set to create individuals with the \textit{weight} attribute that indicates the number of objectives and the type of optimization (maximization/minimization). This is done in the constructor, as follows:
\begin{lstlisting}
	from mggp import mggpElement
	element = mggpElement(weights=(-1,-1))
	element.setPset(maxDelay=1,numberOfVariables=2)
	element.renameArguments({'ARG0':'y1','ARG1':'u1'})
\end{lstlisting}
The argument `\textit{weights=(-1,-1)}' defines an \textit{mggpElement} object that minimizes two objectives. Next, define the cost function to minimize the objectives: (i) the one-step-ahead prediction error and (ii) the number of model terms. The \textit{mggpEvolver} object may be initialized as in the previous example. Finally, use the \textit{runMO()} method to execute the optimization:
\begin{lstlisting}
	def evaluate(ind):
		try:
			element.compile_model(ind)
			theta = element.ls(ind,y,u)
			return element.score_osa(ind, theta, y, u), len(ind)
		except np.linalg.LinAlgError:
			return np.inf, np.inf
	
	hof,log = mggp.runMO(evaluate=evaluate,popPercent=0.8)
\end{lstlisting}
The attribute \textit{popPercent} defines the number of individuals to be kept in the next generation. It is a percentage of \textit{popSize} and selects this number of individuals from the current population plus the resultant offspring observing the Pareto Optimal Set.

\section{Impact}
The \textbf{mggp} is an open source and easy-to-use toolbox that performs the \textit{nonlinear system identification} task via MGGP algorithm. It allows the automatic construction of NARX/NARMAX models using any mathematical function included in the primitive set of an \textit{mggpElement} object (e.g. exponential, sine, tanh, greater than, etc.). The built-in back-shift operator releases the user of the responsibility to predetermine maximum term lags and to build the whole set of candidate terms. Further, researchers can customize the evaluation function using their own algorithms that are not encapsulated in \textit{mggpElement} class. This widens the toolbox application allowing the inclusion of prior knowledge to evaluate individuals and set constraints in the parameters values or in the model structure (gray-box identification). 

\section{Conclusion}
In the present work a new package in \textbf{python} is introduced that performs system identification using NARMAX models. The package encapsulates basic methods of black-box identification such as LS, ELS and FROE and performs mono- and multi-objective optimizations. The users are free to build their cost functions using any method not encapsulated in the package classes. Thus, a broad range of researches in the area of system identification via MGGP is opened. The project is available on GitHub so that users can contribute via
forking and making pull requests, please refer to the repository
indicated in the metadata. 

\subsection{Future versions}
Some issues for future versions are listed in the following.

\textbf{MIMO models:} the extension to multiple-output is essential to widen applications. In its actual version, the \textbf{mggp} package is capable of building up to MISO models.

\textbf{ERR methods:} there are several algorithms of the FROE type  for structure selection using ERR measures with different properties and capabilities. 

\textbf{Human-machine interface:} we have received feedback from users about the use of the package for several inputs. It may get complicated for the users to implement their cost functions and perform analysis. We intend to simplify the interface to ease its use.

\textbf{Optimization algorithms:} there are few built-in operators for cross-over and mutation processes. The algorithms for mono- and multi-optimization are rigid and do not accept modifications. We intend to widen the possibilities taking advantage of the object oriented structure of the package and create an abstract interface for genetic operators. Thus, in future versions, the user may customize the optimization algorithm including genetic operator objects into the \textit{mggpEvolver} object. 

\bibliographystyle{unsrtnat}
\bibliography{references}  

\end{document}